\documentclass[letterpaper, 10 pt, conference]{ieeeconf}  % Comment this line out if you need a4paper

\IEEEoverridecommandlockouts                              % This command is only needed if 
\usepackage{graphics} % for pdf, bitmapped graphics files
\usepackage{epsfig} % for postscript graphics files
\usepackage{color}
\usepackage{mathptmx} % assumes new font selection scheme installed
\usepackage{times} % assumes new font selection scheme installed
\usepackage{amsmath} % assumes amsmath package installed
\usepackage{amssymb}  % assumes amsmath package installed
\usepackage{todonotes}
\usepackage{dirtytalk}
\usepackage{scrextend}
\title{\LARGE \bf
Efficient Computation of Feedback Control for Constrained Systems
}

\author{Forrest Laine$^{1}$, and Claire Tomlin$^{1}$% <-this % stops a space
\thanks{This research is supported by DARPA under the Assured Autonomy Program, 
by NSF under the CPS Frontier VeHICal project, and by the UC-Philippine-California Advanced Research Institute under project IIID-2016-005.}% <-this % stops a space
\thanks{$^{1}$Forrest Laine and Claire Tomlin are with the Department of Electrical Engineering and Computer Sciences, 
        University of California, Berkeley
        {\tt\small forrest.laine@eecs.berkeley.edu}, \tt\small tomlin@eecs.berkeley.edu }
}

\begin{document}

\maketitle
\thispagestyle{empty}
\pagestyle{empty}

%%%%%%%%%%%%%%%%%%%%%%%%%%%%%%%%%%%%%%%%%%%%%%%%%%%%%%%%%%%%%%%%%%%%%%%%%%%%%%%%
\begin{abstract}

A method is presented for solving the discrete-time finite-horizon Linear Quadratic Regulator (LQR) problem subject to auxiliary linear equality constraints, such as fixed end-point constraints. The method explicitly determines an affine relationship between the control and state variables, as in standard Riccati recursion, giving rise to feedback control policies that account for constraints. Since the linearly-constrained LQR problem arises commonly in robotic trajectory optimization, having a method that can efficiently compute these solutions is important. We demonstrate some of the useful properties and interpretations of said control policies, and we compare the computation time of our method against existing methods. 
\end{abstract}

%%%%%%%%%%%%%%%%%%%%%%%%%%%%%%%%%%%%%%%%%%%%%%%%%%%%%%%%%%%%%%%%%%%%%%%%%%%%%%%%
\section{INTRODUCTION} \label{sec:intro}

Due to its mathematical elegance and wide-ranging usefulness, the Linear Quadratic Regulator has become perhaps the most widely studied problem in the field of control theory. Referring to both continuous and discrete-time systems, the LQR problem is that of finding an infinite or finite-length control sequence for a linear dynamical system that is optimal with respect to a quadratic cost function. Either as a stand-alone means for computing trajectories and controllers for linear systems, or as a method for solving successive approximate trajectories for with nonlinear systems, it shows up in one way or another in the computation of nearly all finite-length trajectory optimization problems. 

Because of the importance of trajectory optimization in controlling Robotic systems, and because of the prevalence of the LQR problem in those optimizations, devoting time to highly efficient methods capable of solving LQR-type problems is an important endeavor.  The focus of this paper is on a particular instance of the discrete-time, finite-horizon variant of the LQR problem, being that which is subject to linear constraints. These constrained problems are important in their own-right, and arise in relatively common situations. 

As an example, imagine we want plan a trajectory that minimizes the amount of energy need to get a  robot to some desired configuration. If the dynamics of the robot can be modeled as a linear system, this problem takes the form of linearly-constrained LQR. We can also imagine constraints appearing at multiple stages in the trajectory and having varying dimensions. Perhaps we require that the center of mass of the robot is constrained to not move in the first half of the trajectory. Of course many robots have non-linear dynamics. But even when planning constrained trajectories for non-linear systems, iterative solution methods such as Sequential Quadratic Programming make successive local approximations of the trajectory optimization problem which result in a series of constrained LQR problems to be solved.  We will discuss this relationship in more detail in a later section. 

Understanding that the linearly-constrained LQR problem is common, we provide some context surrounding methods for solving these type of problems. The property that any trajectory must satisfy linear dynamics can be thought of as a sequence of linear constraints on successive states in the trajectory.  And since all auxiliary constraints we consider are also linear, these problems result in quadratic programs (QPs) just as unconstrained LQR problems are QPs \cite{boyd2004convex}. Under standard assumptions, the constrained problems are also strictly-convex and have a unique solution. Unlike unconstrained LQR, however, the presence of additional constraints cause some computational difficulties. 

Looking from a pure optimization standpoint, all of the approaches to solving convex QPs can be applied to the constrained LQR problem without problem. However, using general methods in a naive way fail to exploit the unique structure of the optimal control problem, and suffer a computational complexity which grows cubicly with the time horizon being considered in the control problem (trajectory length). Due to the sparsity of the problem data in the time domain, the KKT conditions of optimality for optimal control problems have a banded nature, and linear algebra packages designed for such systems can be used to solve the problem in a linear complexity with respect to the trajectory length \cite{wright1996applying}. However, these approaches result in what we will call \textit{open-loop} trajectories, producing only numerical values of the state and control vectors making up the trajectory. It is well-known that unconstrained LQR problem offers a solution based on dynamic-programming which is sometimes referred to as the discrete-time Riccati recursion.  This method can also solve unconstrained LQR problems in linear time complexity while \textit{also} providing an affine relationship between the state and control variables. This relationship provides a feedback policy which can be used in control, and offers many advantages over the open-loop variants. 

It is because we would like to derive these policies for the constrained case that the aforementioned computational difficulties show up. The presence of auxiliary constraints have made it such that up until now, a method for the constrained LQR problem analogous to Riccati recursion has not been developed. This is due to the fact that linear constraints of dimension exceeding that of the control can not alway be thought of as time-separable. This means that the choice of control at a particular time-point may not always be able to satisfy a constraint appearing at that time-point (for arbitrary values of the corresponding state at that time). We will see that this complication requires reasoning about future constraints yet to come when computing the control in the present. This is the very reason why, as we will see, existing methods either make restrictive assumptions on the dimension of constraints, or require a higher order of computational complexity to compute solutions.

Because of this, if the problem does not satisfy the restricting assumptions used by existing methods, solution approaches are currently limited to QP solvers and only offer open-loop trajectories, or suffer cubic time-complexity with respect to the trajectory length if control policies are desired. Given this context, we can now state the contribution of this work:

\begin{addmargin}[1em]{2em}
 We present a method for computing constraint-aware feedback control policies for discrete-time, time-varying, linear-dynamical systems which are optimal with respect to a quadratic cost function and subject to auxiliary linear equality constraints. We make no assumptions about the dimension of the constriants. 
 \end{addmargin}

In section \ref{sec:priorwork} we discuss in more detail existing methods which have addressed the same problem and the limitations of those works. In section \ref{sec:method} we formally define the problem and present our method. In section \ref{sec:analysis} we discuss computational complexity, and present an alternative approach to solving the problem. We also demonstrate some of the advantages of the control policies derived from our method when compared to the open-loop solutions, and discuss applicability to SQP methods. 

\section{PRIOR WORK} \label{sec:priorwork}

Consideration of the constrained linear-quadratic optimal control problem extends back to the early days in the field of control. Many authors have presented methods for constraining control systems to a time-invariant linear subspace. The author in \cite{johnson1973stabilization} studied this issue for continuous systems under the name subspace stabilization. In the works \cite{hemami1979modeling} and \cite{yu1996design} the same problem is addressed by designing pole-assignment controllers. More recently, \cite{posa2016optimization} utilize a very similar method to generate a time-varying controller for tracking existing trajectories. This method is also derived in continuous-time, and hence requires the constraint-dimension to be constant. 
The authors in \cite{ko2007optimal} developed a more comprehensive method for computing optimal control policies for discrete-time, time-varying objective functions, but only considers a single time-invariant constraint of constant dimension. In \cite{park2008lq} a method is presented for solving continuous- and discrete-time LQR problems with fixed terminal states. This method is able to reason about a constraint only appearing at a portion of the trajectory, being the end, but does not account for additional constraints appearing at other times, however. 
Perhaps the most general method for computing linearly constrained LQR control policies was presented in \cite{sideris2011riccati}. However, that method suffers a computational complexity which scales cubicly in the worst-case, i.e. when many constraints which have dimension exceeding the control dimension are present. As a part of the method presented in \cite{xie2017differential}, a technique for satisfying linear constraints at arbitrary times in the trajectory is presented, but that method assumes that the constraint dimension does not exceed that of the control. Most recently, \cite{giftthaler2017projection} present a method for solving problems with time-varying constraints, but still require that the relative-degree of these constraints does not exceed 1. This is a slightly less-restrictive condition than requiring the dimension of the constraints be less than that of the control, but still limits the applicability of that method in that it can not handle full-state constraints when the control dimension is less than that of the state dimension. 
As mentioned above, the problem can also be solved using numerical linear algebra techniques, as discussed for example in \cite{wright1999numerical} and particularly for the optimal control problem, \cite{wright1996applying}. Again, these methods are very general and efficient but fail to produce the desired feedback control policies.  

The method we present combines the desirable properties of all these methods into one. The contribution of this method is that it is capable of generating optimal feedback control policies for general, discrete-time, linearly-constrained LQR problems while maintaining a linear computational complexity with respect to control horizon.  To the best of our knowledge, the approach we present is the only method in existence that is capable of this.  \nocite{goebel2005continuous}  \nocite{mare2007solution} \nocite{cannon2008efficient} \nocite{scokaert1998constrained}

\section{PROBLEM AND METHOD} \label{sec:method}

The method we present here is a means of deriving optimal feedback control policies for the following problem:

\begin{subequations}
\begin{align}
    & \min_{x_0, u_0,...,u_{T-1}, x_T} cost_T(x_T) + \sum_{t=0}^{T-1} cost_t(x_t, u_t) \label{obj:globalobjwords} \\
    & \text{s.t.} \ \ \ \ dynamics_t(x_{t+1}, x_t, u_t) = 0 \ \forall t \in \{0,...,T-1\} \label{const:dynamicswords} \\
    & \ \ \ \ \ \ \ \  x_0 = x_{init} \label{eq:init} \\ 
    & \ \ \ \ \ \ \ \  constraint_t(x_t, u_t) = 0, \ \forall t \in \{0,...,T-1\} \label{const:twords}\\
    & \ \ \ \ \ \ \ \  constraint_T(x_T) = 0 \label{const:Twords}
\end{align} \label{opt:globalwords}
\end{subequations}
    % & \text{s.t.} \ \ \ x_{t+1} = F_{x_t} x_t + F_{u_t} u_t + f_{1_t}, \ \forall t \in \{0,...,T-1\} \\
    %& \text{s.t.} \ \ \ x_{t+1} = F_{x_t} x_t + F_{u_t} u_t + f_{1_t}, \ \forall t \in \{0,...,T-1\} \\
    %& \ \ \ \ \ \ \ \ \ x_0 = x_{init} \\ 
    %& \ \ \ \ \ \ \ \ \ 0 = G_{x_t} x_t + G_{u_t} u_t + g_{1_t}, \ \forall t \in \{0,...,T-1\} \label{const:lin}\\
    %& \ \ \ \ \ \ \ \ \ 0 = G_{x_T} x_T + g_{1_T} \label{const:lin_end}
Where $x_t \in \mathbb{R}^n$, $u_t \in \mathbb{R}^m$, and the functions 
\begin{align*}
&cost_t: \mathbb{R}^n \times \mathbb{R}^m \to \mathbb{R} & &cost_T: \mathbb{R}^n \to \mathbb{R} \\
&constraint_t: \mathbb{R}^n \times \mathbb{R}^m \to \mathbb{R}^{l_t} & &constraint_T: \mathbb{R}^n \to \mathbb{R}^{l_T} \\
&dynamics_t : \mathbb{R}^n \times \mathbb{R}^n \times \mathbb{R}^m \to \mathbb{R}^n
\end{align*} are defined as:
\begin{align}
    &cost_t(x, u) = \frac{1}{2} \begin{pmatrix} 1 \\ x \\ u \end{pmatrix}^\intercal \begin{pmatrix} 0 & q_{x1_t}^\intercal & q_{u1_t}^\intercal \\ q_{x1_t} & Q_{xx_t} & Q_{ux_t}^\intercal \\ q_{u1_t} & Q_{ux_t} & Q_{uu_t} \end{pmatrix}\begin{pmatrix} 1 \\ x \\ u \end{pmatrix} \label{eq:costtform} \\
    &cost_T(x) = \frac{1}{2} \begin{pmatrix} 1 \\ x \end{pmatrix}^\intercal \begin{pmatrix} 0 & q_{x1_T}^\intercal \\ q_{x1_T} & Q_{xx_T} \end{pmatrix}\begin{pmatrix} 1 \\ x \end{pmatrix} \\
    &dynamics_t(x_{t+1}, x_t, u_t) = x_{t+1} - (F_{x_t} x_t + F_{u_t} u_t + f_{1_t}) \label{eq:dynamicsform}\\
    &constraint_t(x_t, u_t) = G_{x_t} x_t + G_{u_t} u_t + g_{1_t} \label{eq:constrainttform} \\
    &constraint_T(x_T) = G_{x_T} x_T + g_{1_T} \label{eq:constraintTform},
\end{align}
where $l_t$ (for $0 \leq t < T$) and $l_T$ are the dimensions of the constraints at the corresponding times. 

In the above expressions, and in the rest of this paper, coefficients are assumed to have dimension such that the expression makes sense. We assume for now that the coefficients $Q_{uu_t}$ of the quadratic functions $cost_t$ are positive-definite, and that  $Q_{xx_t} - Q_{ux_t} Q_{uu_t}^{-1}  Q_{ux_t}^\intercal $ is positive semi-definite. This assumption is possible to be relaxed, and we will discuss this below.  

%We also assume that the  constraints (\ref{const:dynamicswords}-\ref{const:Twords}) are linearly independent. 
%We also make use of the following notation: coefficients appearing in linear expressions, such as in (\ref{eq:dynamicsform}), have the form $C_{v_t}$ for matrix coefficients and $c_{1_t}$ for vector coefficients, where $v$ is either '$x$' or '$u$', representing the coefficient multiplying $x_t$ or $u_t$. We use the subscript '$1_t$' to mark the affine coefficient of an expression. For quadratic expressions, such as (\ref{eq:costtform}), matrix coefficients have the form $C_{vw_t}$ and vector coefficients have the form $c_{v1_t}$. In the matrix terms, $v$ and $w$ are either '$x$' or '$u$' and represent the two variables being multiplied and scaled by the coefficient. In the vector coefficient terms $v$ indicates the variable with which the dot product is taken. 

\subsection*{Constrained LQR}

The method for computing the constrained control policies will follow a dynamic programming approach. Starting from the end of the trajectory and working towards the beginning, a given control input $u_t$ will be chosen such that for any value of the resulting state $x_t$, the control will satisfy all constraints imposed at time $t$, as well as any constraints remaining to be satisfied in the remainder of the trajectory, if possible. 

If there are degrees of freedom in the control input that do not affect the constraint, the portion of the control lying in the null-space of the constraint will be chosen such as to minimize the cost in the remainder of the trajectory. If the constraint is unable to be satisfied by the control for arbitrary states, the control will minimize the sum of squared residuals of the constraints. This has the effect of eliminating $r$ dimensions of the constraint, where $r$ is the rank of the constraint coefficient multiplying $u_t$. For a trajectory to satisfy the constraint in this case, the state $x_t$ must therefore be such that the constraint residuals will be zero. This can be enforced by passing on a residual linear constraint to the choice of control at the preceding time, $u_{t-1}$ (and controls preceding that, if necessary). 

To formalize this procedure, we introduce a time-varying quadratic function, $cost\_to\_go_t: \mathbb{R}^n \to \mathbb{R}$, representing the minimum possible cost remaining in the trajectory from stage $t$ onward as a function of state. Additionally, we introduce a linear function $constraint\_to\_go_t : \mathbb{R}^n \to \mathbb{R}^{p_t}$, which defines through a constraint on $x_{t}$  the subspace of admissible states such that the control $u_t$ will be able to satisfy the constraints in the remainder of the trajectory.  Here $p_t$ is the dimension of constraints needed to enforce this condition.  These functions are defined as follows:
\begin{align}
    cost\_to\_go_t(x) &= \frac{1}{2}
     \begin{pmatrix} 1 \\ x \end{pmatrix}^\intercal \begin{pmatrix} 0 & v_{x1_t}^\intercal \\ v_{x1_t} & V_{xx_t} \end{pmatrix}\begin{pmatrix} 1 \\ x \end{pmatrix} \label{eq:costtogo} \\
     constraint\_to\_go_t(x) &= H_{x_t} x + h_{1_t} . \label{eq:constrainttogo}
\end{align}
We initialize these terms at time $T$: 
\begin{equation}
\begin{aligned}
    V_{xx_T} &= Q_{xx_T} & v_{x1_T} &= q_{x1_T} \\
    H_{x_T} &= G_{x_T}  & h_{1_T} &= g_{1_T}.
\end{aligned}
\end{equation}
Note that in the value function (\ref{eq:costtogo}) we do not include any constant terms (which would appear in the $0$ block of (\ref{eq:costtogo})). This is because the calculations we will derive do not depend on them, and so we omit them for clarity. 

Given the above definitions, starting at $T-1$ and working backwards to $0$, we solve the following optimization problem for each time $t$:
\begin{subequations} \begin{align}
    u_t^*(x_t) = \ &\text{arg}\min_{u_t} \ cost_t(x_t, u_t) + cost\_to\_go_{t+1}(x_{t+1}) \\
    \text{s.t.} \ \ \ & 0 = dynamics_t(x_{t+1}, x_t, u_t)  \label{eq:dyndp} \\
    & u_t \in \text{arg}\min_{u} \| \begin{bmatrix} constraint_t(x_t, u) \\ constraint\_to\_go_{t+1}(x_{t+1}) \end{bmatrix} \|_2  \label{eq:setdp}
\end{align}  \label{opt:words} \
\end{subequations}
To see why we want to solve this problem, we first simplify it by using the form of (\ref{eq:dyndp}) to eliminate $x_{t+1}$, and plug in coefficients:
\begin{subequations}
\begin{align} \label{eq:simple}
    u_t^*(x_t) &= \text{arg}\min_{u_t}  \frac{1}{2} \begin{pmatrix} 1 \\ x_t \\ u_t \end{pmatrix}^\intercal \begin{pmatrix} 0 & m_{x1_t}^\intercal  & m_{u1_t}^\intercal \\ m_{x1_t} & M_{xx_t} & M_{ux_t}^\intercal \\ m_{u1_t} & M_{ux_t} & M_{uu_t} \end{pmatrix}\begin{pmatrix} 1 \\ x_t \\ u_t \end{pmatrix}  \\
    &\text{s.t.} \ \ u_t \in \text{arg}\min_u  \| N_{x_t} x_t + N_{u_t} u + n_{1_t} \|_2 .\label{const:simple}
\end{align} \label{opt:simple}
\end{subequations}
Where the above terms are defined as:
\begin{equation}
\begin{aligned}
    m_{x1_t} &= q_{x1_t} + F_{x_t}^\intercal v_{x1_{t+1}} & m_{u1_t} &= q_{u1_t} + F_{u_t}^\intercal v_{x1_{t+1}} \\ 
    M_{xx_t} &= Q_{xx_t} + F_{x_t}^\intercal V_{xx_{t+1}} F_{x_t} &  M_{uu_t} &= Q_{uu_t} + F_{u_t}^\intercal V_{xx_{t+1}} F_{u_t} \\
    M_{ux_t} &= Q_{ux_t} + F_{u_t}^\intercal V_{xx_{t+1}} F_{x_t}  &  N_{x_t} &= \begin{pmatrix} G_{x_t} \\ H_{x_{t+1}}F_{x_t} \end{pmatrix} \\ 
    N_{u_t} &= \begin{pmatrix} G_{u_t} \\ H_{x_{t+1}} F_{u_t} \end{pmatrix} & n_{1_t} &= \begin{pmatrix} g_{1_t} \\ H_{x_{t+1}}f_{1_t} + h_{1_{t+1}} \end{pmatrix}.
\end{aligned}
\end{equation}

%If the state $x_t$ is such that $N_{x_t} x_t + n_{1_t}$ is in the range-space of $N_{u_t}$, then the we will have that $constraint_t(x_t,u_t)=0$ and $constraint\_to\_go_{t+1}(x_{t+1})=0$. Since we have assumed that if these constraints are in fact $0$, then the remaining trajectory from time $t$ to $T$ will be feasible. However, if $N_{x_t} x_t + n_{1_t}$ is not in the range-space of $N_{u_t}$, solving (\ref{opt:words}) with those constraints constrained to be zero results in an infeasible problem. By choosing $u_t$ as in (\ref{eq:setdp}), the control is chosen such as to satisfy as much of the constraint as possible. 
%We also define the useful value 
%\begin{equation}
%r_t = rank(N_{u_t}).
%\end{equation}
% We have by the definition of $N_{x_t}$, $N_{u_t}$, and $n_{1_t}$ that if (\ref{const:simple}) is satisfied, then all constraints are satisfied for the trajectory from stages $t$ to $T$. We have already assumed that a solution to the full optimization problem (\ref{eq:obj}) exists. Therefore, it follows that the $x_t$ that enters in (\ref{eq:simple}-\ref{const:simple}) is such that a solution to the simplified problem also exists. 
%This will always be the case if the number of linearly-independent constraints exceeds the dimension of the control vector, such as when $n > m$ and a full-state constraint is imposed, and there is no way  More generally, this will happen when the matrix $N_{u_t}$ is not full row-rank. 

If, for a particular $x_t$, the vector $N_{x_t}x_t + n_{1_t}$ is not in the range-space of $N_{u_t}$, then the minimizer of (\ref{const:simple}) will have a non-zero constraint residual. This is why we can not enforce the conditions $constraint_t(x_t)=0$ and $constraint\_to\_go_{t+1}(x_{t+1})=0$ in (\ref{opt:words}), since that would result in a infeasible problem in general. When $x_t$ is such that the constraints are not feasible, this implies that even with the best effort done by the control to satisfy the constraints, the state $x_t$ can not be arbitrary. We must therefore place the requirement on $x_t$ that $N_{x_t}x_t + n_{1_t}$ does in fact lie in the range-space of $N_{u_t}$. This constraint will then be passed on to the choice of control at the preceding time. We will derive explicitly what this constraint on $x_t$ looks like after determining a closed-form solution of $u_t^*(x_t)$. To do this, we first write an equivalent problem to (\ref{opt:simple}):
\begin{subequations}
\begin{align}
&\begin{aligned}
y_t^*, w_t^* = \text{arg}\min_{y_t, w_t}  \frac{1}{2} \| N_{x_t} x_t + N_{u_t} P_{y_t} y_t + n_{1_t} \|_2  + \ \ \ \ \ \ \
  \\  \frac{1}{2} \begin{pmatrix} 1 \\ x_t \\ Z_{w_t} w_t \end{pmatrix}^\intercal \begin{pmatrix} 0 & m_{x1_t}^\intercal  & m_{u1_t}^\intercal \\ m_{x1_t} & M_{xx_t} & M_{ux_t}^\intercal \\ m_{u1_t} & M_{ux_t} & M_{uu_t} \end{pmatrix}\begin{pmatrix} 1 \\ x_t \\ Z_{w_t} w_t \end{pmatrix} \label{eq:vwsoln} \end{aligned} \\
& u_t^* = P_{y_t} y_t^* + Z_{w_t} w_t^* \label{eq:udirectsum}
\end{align} \label{opt:infeasible}
\end{subequations}
Here, $Z_{w_t}$ is chosen such that the columns form an ortho-normal basis for the null-space of $N_{u_t}$, and $P_{y_t}$ is chosen such that its columns form a ortho-normal basis for the range-space of $N_{u_t}^\intercal$.  Hence $N_{u_t}$ and $P_{y_t}$ are also orthogonal and their columns together span $\mathbb{R}^m$. Because of the orthogonality of these two sub-spaces, any control signal $u_t$ corresponds to a unique $y_t$ and $w_t$ \cite{callier2012linear}. 

%There are a few subtleties that are important to note here. Although the choice of $P_{y_t}$ and $Z_{w_t}$ are not unique, the values $P_{y_t}y_t^*$ and $Z_{w_t}w_t^*$ are unique. This is because we have assumed $Q_{uu_t}$ is positive definite, $Q_{xx_t} - Q_{ux_t}Q_{uu_t}^{-1}Q_{ux_t}^\intercal$ is positive semi-definite\footnote{These conditions are actually over-restrictive. We instead only require that $Z_{w_t}^\intercal M_{uu_t} Z_{w_t}$ is positive-definite, although this condition is hard to verify a-priori for general problems where the stated conditions are not met.}, and because the matrix $N_{u_t}P_{y_t}$ is full-rank by construction. Therefore the problem (\ref{opt:infeasible}) is strictly convex with respect to both $y_t$ and $u_t$. Since $P_{y_t}$ and $Z_{w_t}$ are both ortho-normal, the minimum of (\ref{opt:infeasible}) is unchanged for different choice of these matrices. Therefore, the vectors $P_{y_t}y_t^*$ and $Z_{w_t}w_t^*$ must be unique for any appropriate choice of $P_{y_t}$ and $Z_{w_t}$. 

%Since the constraints imposed on the trajectory in the interval $\{t,...,T\}$ can only be satisfied if the constraint residual in (\ref{const:simple} and \ref{eq:vwsoln}) is zero, we enforce that this must be the case by putting a constraint on the state $x_t$.  The form of this constraint will become clear as we derive the relationship between the optimal control signal $u_t^*$ as a function of the state $x_t$.  
The solution to (\ref{opt:infeasible}), which is an unconstrained problem, is:
\begin{align}
	y_t^* &= -(N_{u_t}P_{y_t})^{\dagger} (N_{x_t} x_t + n_{1_t}) \label{eq:controlvupdate} \\
	w_t^* &= -(Z_{w_t}^\intercal M_{uu_t} Z_{w_t})^{-1} Z_{w_t}^\intercal (M_{ux_t} x_t + m_{u1_t}). \label{eq:controlwupdate} 
\end{align}
 
 In the case that $P_{y_t}$ is a zero matrix (i.e. $\text{dim}(\text{null}(N_{u_t}))=m$), $Z_{w_t} = I_m$ (Identity matrix $\in \mathbb{R}^{m\times m}$), and $y_t$ has dimension 0. Correspondingly, when the nullity of $Z_{w_t}$ is 0, $P_{y_t} = I_m$ and $w_t$ has dimension 0. Therefore, in these cases, we ignore the update (\ref{eq:controlvupdate} or \ref{eq:controlwupdate}) that is of size $0$.  With this in mind, and combining terms, we can express the control $u_t$ in closed-form as an affine function of the state $x_t$:
 \begin{align}
 	u_t^* &= K_{x_t} x_t + k_{1_t}  \label{eq:controlpolicy} \\
	K_{x_t} &= -\big( P_{y_t}(N_{u_t}P_{y_t})^\dagger N_{x_t} + Z_{w_t} (Z_{w_t}^\intercal M_{uu_t} Z_{w_t})^{-1}Z_{w_t}^\intercal M_{ux_t} \big) \label{eq:K} \\
	k_{1_t} &= -\big( P_{y_t}(N_{u_t}P_{y_t})^\dagger n_{1_t} + Z_{w_t} (Z_{w_t}^\intercal M_{uu_t} Z_{w_t})^{-1}Z_{w_t}^\intercal m_{u1_t} \big) \label{eq:k}
 \end{align}
 
Since the control is a function of the state, we can also express the constraint residual as a function of the state. We can let the function $constraint\_to\_go_t$ to be the value of the constraint residual (\ref{eq:setdp}). We substitute  (\ref{eq:controlpolicy}, \ref{eq:K}, \ref{eq:k}) into  (\ref{const:simple}) to obtain:
\begin{align}
\begin{aligned}
	constraint\_to\_go_t(x_t) &= \\
	N_{x_t}x_t - N_{u_t}P_{y_t}&(N_{u_t}P_{y_t})^\dagger(N_{x_t}x_t +n_{1_t}) + n_{1_t} \label{eq:constrainttogo}
\end{aligned}
\end{align}
This results in the update for the terms $H_{x_t}$ and $h_{1_t}$:
\begin{align}
H_{x_t} &= (I - N_{u_t}P_{y_t}(N_{u_t}P_{y_t})^\dagger)N_{x_t} \\
h_{1_t} &= (I - N_{u_t}P_{y_t}(N_{u_t}P_{y_t})^\dagger)n_{1_t}.
\end{align}	

Here $I$ is the identity matrix having the same leading dimension as $N_{x_t}$. By observing these updates, we see that the terms in (\ref{eq:constrainttogo}) are computed by projecting $N_{x_t} x_t + n_{1_t}$ into the kernel of $(N_{u_t}P_{y_t})^\intercal$, as we would expect given the discussion above.  Hence, the residual constraint will lie in a subspace of dimension no larger than the nullity of $(N_{u_t}P_{y_t})^\intercal$. We can therefore remove redundant constraints by removing linearly-dependent rows of the matrix $\begin{bmatrix} h_{1_t} & H_{x_t} \end{bmatrix}$, in order to maintain a minimal representation and keep computations small. 

Note that if the matrix $\begin{bmatrix} h_{1_t} & H_{x_t} \end{bmatrix}$ is full column rank at any time $t$, then there exists no $x_t$ that can satisfy the constraints, and we have detected that the trajectory optimization problem (\ref{opt:globalwords}) is infeasible. 
 
We now plug the expression for the control in to the objective function of our optimization problem (\ref{eq:simple}) to obtain an update on our value function as a function of the state (again, omitting constant terms):
 \begin{align}
    cost\_to\_go_t(x_t) &= \frac{1}{2} \begin{pmatrix} 1 \\x_t\\ u_t^* \end{pmatrix} ^\intercal  
    \begin{pmatrix} 0 & m_{x1_t}^\intercal  & m_{u1_t}^\intercal \\ m_{x1_t} & M_{xx_t} & M_{ux_t}^\intercal \\ m_{u1_t} & M_{ux_t} & M_{uu_t}\end{pmatrix}
    \begin{pmatrix} 1 \\x_t\\ u_t^* \end{pmatrix} \\
    &= \frac{1}{2} \begin{pmatrix} 1 \\x_t \end{pmatrix} ^\intercal  
    \begin{pmatrix} 0 & v_{x1_t}^\intercal \\ v_{x1_t} & V_{xx_t} \end{pmatrix}
    \begin{pmatrix} 1 \\x_t \end{pmatrix},
\end{align}

where terms are defined as
\begin{align}
    V_{xx_t} &=  M_{xx_t} + 2 M_{ux_t}^\intercal K_{x_t} + K_{x_t}^\intercal M_{uu_t} K_{x_t}  \label{update:V} \\
    v_{x1_t} &= m_{x1_t} + K_{x_t}^\intercal m_{u1_t} + (M_{ux_t}^\intercal + K_{x_t}^\intercal M_{uu_t}) k_{1_t} \label{update:v}.
\end{align}

We have now presented updates for the terms $V_{xx_t}$, $v_{x1_t}$, $H_{x_t}$, and $h_{1_t}$, and computed control policy terms $K_{x_t}$ and $k_{1_t}$ in the process. The sequence of control policies $\{K_{x_t}, k_{1_t}\}_{t\in\{0,...,T-1\}}$ will produce a sequence of states and controls that are optimal for our original problem (\ref{opt:globalwords}). 

\section{Analysis} \label{sec:analysis}
In the preceding section, we have presented a method for computing control policies for the equality-constrained LQR problem (\ref{opt:globalwords}). In this section we will analyze the method and the resulting policies by evaluating the computational complexity of the method and by relating the policies to those that are produced in standard, unconstrained LQR.

\subsection{Computation} \label{sec:computation}

\begin{table}[htb]
\centering
\begin{tabular}{|c|c|c|c|c|c|}
\hline
n                       & m                      & T                        & \multicolumn{1}{l|}{\% Constrained} & LAPACK (s) & CLQR (s)\\ \hline
40                      & 10                     & 250                      & 0                                   & 0.089      & 0.031           \\ \hline
40                      & 10                     & 250                      & 90                                  & 0.095      & 0.040           \\ \hline
40                      & 10                     & 125                      & 90                                  & 0.046      & 0.021           \\ \hline
9                       & 2                      & 250                      & 0                                   & 0.002      & 0.004           \\ \hline
\multicolumn{1}{|l|}{9} & \multicolumn{1}{l|}{2} & \multicolumn{1}{l|}{250} & 50                                  & 0.002      & 0.007           \\ \hline
\multicolumn{1}{|l|}{9} & \multicolumn{1}{l|}{2} & \multicolumn{1}{l|}{125} & 50                                  & 0.001      & 0.003           \\ \hline
\end{tabular}
\caption{Comparing computation times of constrained and unconstrained LQR problems between our constrained LQR method (CLQR) and a method using LAPACK to directly solve the KKT system of equations.}
\label{table:tab}
\end{table}

We mentioned that one of the contributions of this method is its computational efficiency compared to existing results. Due to the dynamic-programming nature of this method, the computational time-dependence on trajectory length is linear, irrespective of the dimension of auxiliary constraints. 

In each iteration of the dynamic programming backups, the heavy computations involve computing the null- and range-space representations of $N_{u_t}$ and $N_{u_t}^\intercal$, respectively, and then computing the pseudo-inverse of $N_{u_t}P_{y_t}$. These operations can all be done by making use of one singular value decomposition (SVD). The dimension of $N_{u_t}$ is no greater than $(2n+m) \times m$ where $n$ is the dimension of the state and $m$ is the dimension of the control signal. Computation complexity of the SVD is thus $O((2n+m)^2m + m^3)$ \cite{Golub89a}. We also make use of a decomposition on the terms $\begin{bmatrix} h_{1_t} &  H_{x_t}\end{bmatrix}$ to remove redundant constraints, which requires computations on the order of $O(n^3)$. The remaining computations are numerous matrix-matrix products and matrix-inversions with terms having dimension no larger than $2n+m \times n$. Thus, the overall order of the method presented here is $O(T(\kappa_1n^3 + \kappa_2n^2m + \kappa_3m^2n + \kappa_4m^3))$, where $\kappa_1,\kappa_2,\kappa_3,\kappa_4$ are some positive scalars. 

Therefore, the method we present here has computational complexity which is roughly equivalent to known solutions based on using a banded-matrix solver on the system of KKT conditions \cite{demmel1997applied} \cite{wright1993interior}. This is not surprising, since our method can be thought of as performing a specialized block-substitution method on the system of KKT conditions, and hence a specialized block-substitution solver for the particular structure arising in constrained optimal control problems. 

In table \ref{table:tab} we show a comparison of computation times between our method and the method 'DGBTRS' from the well-known linear algebra package LAPACK \cite{laug}. All times are taken as the minimum over 10 trials, run on a 2013 MacBook Air with a 2-core GHz Intel Core i7 processor. The LAPACK method performs gaussian elimination on the banded KKT system of equations. We make this comparison for varying problem sizes and percentage of the number of independent constraints relative to the total number degrees of freedom in the problem. For problems of relatively small size, we see that LAPACK offers superior speed, even in the standard unconstrained LQR case. However, as the problem size grows, we see that our method quickly becomes more efficient than the LAPACK solution. 

\subsection{Infinite Penalty Perspective}
We here consider an alternative way to solve (\ref{opt:globalwords}), being the quadratic penalty approach. It is known that we can solve equality-constrained quadratic programs by solving an unconstrained problem, where the linear constraint terms are penalized in the objective as an infinitely weighted cost on the sum of squared constraint residuals \cite{bertsekas1999nonlinear}. Therefore, in light of our original problem (\ref{opt:globalwords}), we could penalize the constraints (\ref{const:twords} and \ref{const:Twords}) in this way, which would result in a standard (from a structural standpoint) LQR problem, where some of the cost terms are weighted infinitely high. The resulting problem would appear as 
\begin{subequations}
\begin{align}
& \min_{u_0,...,u_{T-1}} constraint\_penalized\_cost_T(x_T) \  + 
\\ & \ \ \ \ \ \ \ \ \ \  \sum_{t=0}^{T-1} constraint\_penalized\_cost_t(x_t, u_t) \\
    & \text{s.t.} \ \ \ dynamics_t(x_{t+1}, x_t, u_t) = 0 \ \forall t \in \{0,...,T-1\} \\
    & \ \ \ \ \ \ \  x_0 = x_{init} .
\end{align} \label{opt:quadpenalty}
\end{subequations}
Here the modified cost functions are defined as
\begin{align}
&\begin{aligned}
constraint\_modified\_cost_t(x, u) = \ \ \ \ \ \ \ \ \ \ \ \ \ \ \ \ \ \ \ \   \\
 cost_t(x, u) + \frac{1}{\epsilon} \| constraint_t(x,u)\|^2_2 
 \end{aligned} \\
 &\begin{aligned}
constraint\_modified\_cost_T(x) = \ \ \ \ \ \ \ \ \ \ \ \ \ \ \ \ \ \ \ \ \ \  \\
 \ \ \ \ \ \ cost_T(x) +  \frac{1}{\epsilon} \|constraint_T\|^2_2.
 \end{aligned}
\end{align}
In practice, we can not penalize the constraint terms by infinity (by letting $\epsilon \to 0_+$), but it may suffice to penalize the constraints by some very large constant. Because the optimal control problems we typically solve are based on approximate models of systems, solving an 'approximately' constrained system may be adequate, since super-high precision on the solution is overkill. In these cases, one could consider solving the unconstrained penalized problem (\ref{opt:quadpenalty}). The necessary computation for computing its solution is slightly less than the approach developed in section \ref{sec:method}, as was shown in Table \ref{table:tab}. 
\begin{figure}[htb] 
  \centering
  \includegraphics[scale=0.18]{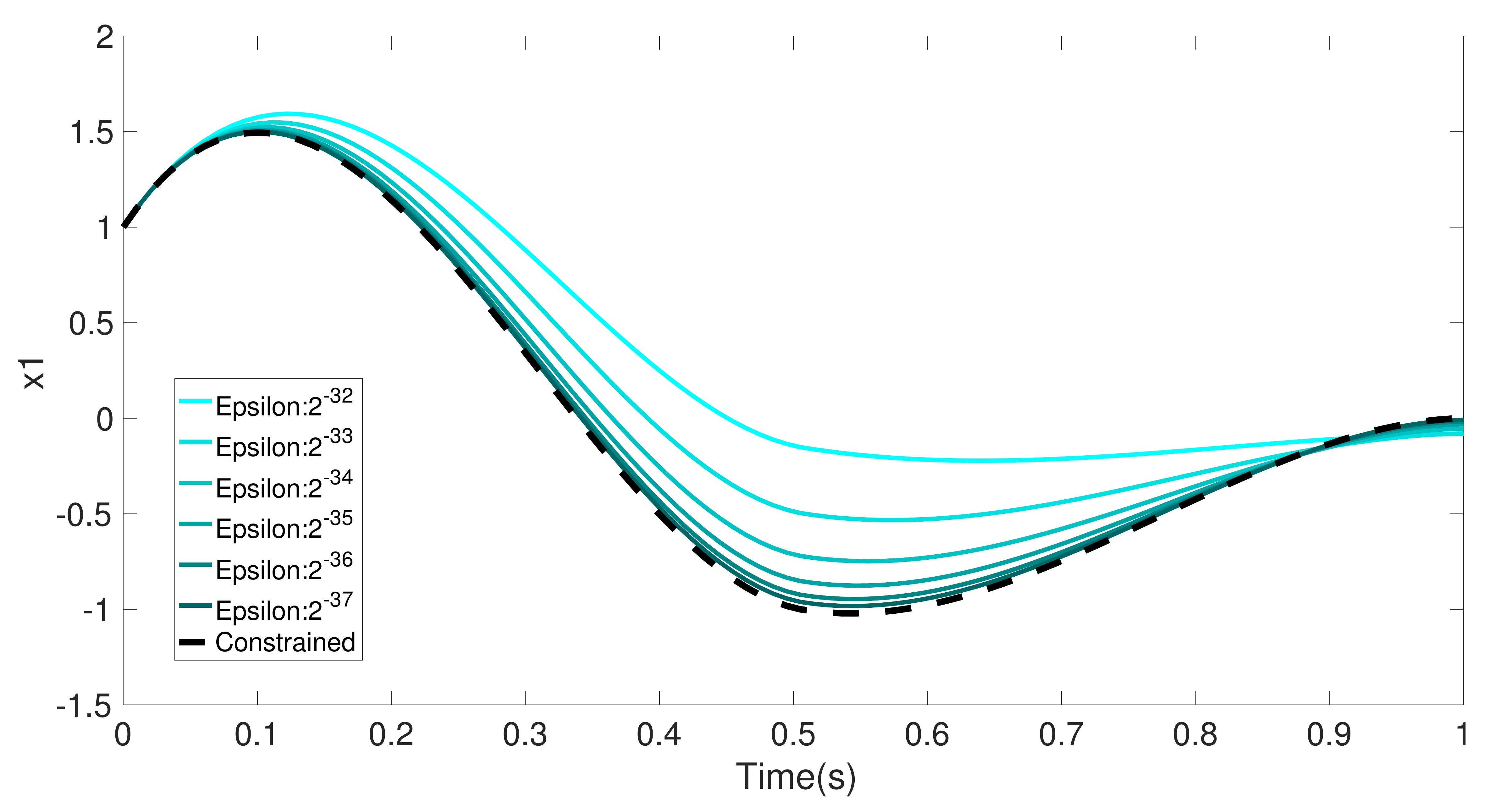} 
  \caption{Limiting behavior of the penalty approach for problem (\ref{opt:doubleint})}
  \label{figure:doubleint}
\end{figure}

We illustrate this relationship between the two methods using a very simple example. Consider the constrained LQR problem for a discrete-time double integrator below:
\begin{subequations}
\begin{align}
	\min_{u_0,...,u_{T-1}}& \  \sum_{t=0}^{T-1} \|u_t\|_2^2 \\
	\text{s.t.} \ \ \ x_{t+1} &= \begin{bmatrix} 1 & dt \\ 0 & 1 \end{bmatrix} x_t + \begin{bmatrix} 0 \\ dt \end{bmatrix} u_t  \\
	x_0 &= \begin{bmatrix} 1& 1\end{bmatrix} ^\intercal \\
	x_{T/2} &=  \begin{bmatrix} -1& -1\end{bmatrix} ^\intercal \\
	x_T &= \begin{bmatrix} 0 & 0 \end{bmatrix} 
\end{align} \label{opt:doubleint}
\end{subequations}
For this example, we let $dt=0.01$ and $T=100$ to simulate a one second trajectory. In figure \ref{figure:doubleint}, trajectories of the first element of $x_t$ can be seen for the solution to the explicitly constrained formulation as well as solutions computed using the penalty formulation (\ref{opt:quadpenalty}) for varying values of $\epsilon$. As can be seen, as $\epsilon \to 0_+$, the solutions of the penalty method converge to that of the explicitly constrained method. %We note that very small values of $\epsilon$ (and hence very high values of $1/\epsilon$) are required to achieve comparable results on this small problem.

While this simpler approach might seem an enticing alternative to the approach outlined in section \ref{sec:method}, we maintain that our method which handles constraints explicitly is still important. Our method ensures the optimal solution without guessing a sufficient value of $\epsilon$. In applications where correct solutions are needed, such as using this method in the context of an SQP approach (discussed more below), iteratively updating the penalty parameter until acceptable constraint satisfaction might be much slower than computing the analytic solution from the start. 

%Furthermore, using our method, constrained LQR problems can be solved explicitly as a problem parameterized by variable constraint values. Consider the constrained LQR example above (\ref{opt:doubleint}). Using our method, we can solve for this policy explicitly without knowing the endpoint, which would result in control policies that have a linear dependence on the unknown endpoint. We can then reuse the policy to get to arbitrary terminal states by plugging in said endpoint. This idea, and extensions resulting in parallelization of computation are investigated further in another paper \cite{parallel paper}. 

\subsection{Disturbance Rejection} 

\begin{figure}[htb] 
  \centering
  \includegraphics[scale=0.18]{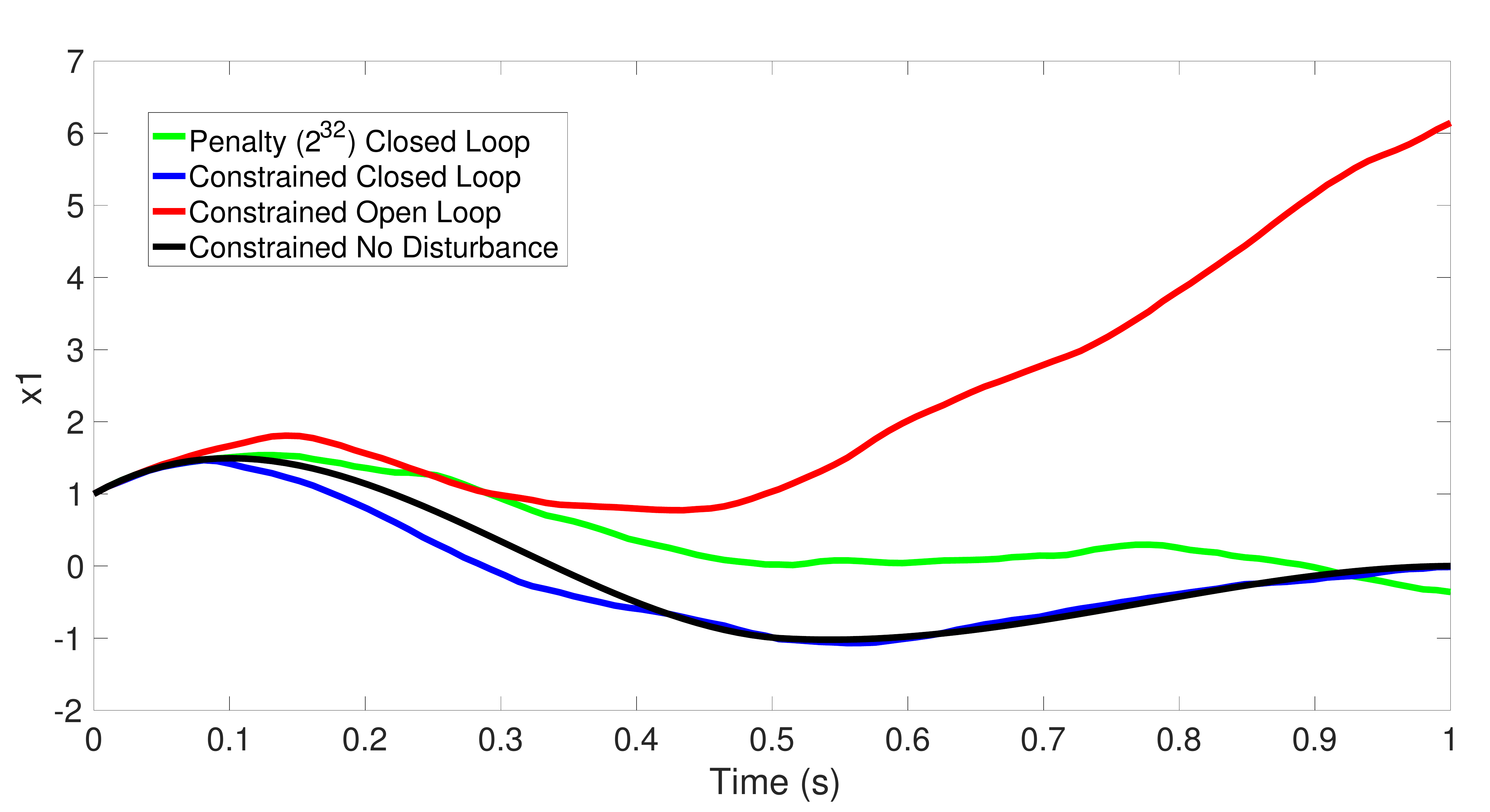} 
  \caption{Robust constraint satisfaction for problem (\ref{opt:doubleint}) subject to additive input noise. }
  \label{figure:disturbed}
\end{figure}

Another benefit of the control policies we have generated is in robustly satisfying constraints. Consider again the example (\ref{opt:doubleint}). Let us compare the performance of executing the open-loop control signal as would be generated when using a gaussian elimination technique as discussed above, compared to executing the constrained feedback policies, in the presence of unforeseen disturbances. In figure \ref{figure:disturbed}, we see the comparison of the open loop control policy compared to the feedback policy when executed on a 'true' system with dynamics when the input $u_t$ is corrupted by gaussian noise. 
%\begin{align} 
%x_{t+1} =  \begin{bmatrix} 1 & dt \\ 0 & 1 \end{bmatrix} x_t + \begin{bmatrix} 0 \\ dt \end{bmatrix} u_t + \begin{bmatrix} 0 \\ \mathcal{N}(0,1) \end{bmatrix}. \label{eq:disturbed}
%\end{align}
%Here $\mathcal{N}(0,1)$ is a random disturbance drawn from a standard normal distribution. Note that given these dynamics, the optimal control signal is the same as if computed assuming the dynamics in (\ref{opt:doubleint}) \cite{kwakernaak1972linear}). 
We see (as would be expected) that the open loop signal strays far from satisfying either of the equality constraints, where as by using the constrained feedback policies, they are still nearly satisfied. This is a purely empirical argument, but demonstrates a simple case in which the benefits of the generated control policies are seen. More in-depth analysis of the robustness properties of constraint-aware feedback policies can be seen in \cite{ko2007optimal} for a time-invariant constraint, and a similar analysis could be done for the general constraint policies presented here, but is left for future work. 

\subsection{Application to Sequential Quadratic Programming}
Due to the generality and computational efficiency of our method, we have mentioned that it is well-suited for algorithms for solving more complicated optimal control problems. In particular, consider the more general version of problem (\ref{opt:globalwords}) where the cost functions might be non-quadratic or even non-convex, and the dynamic and auxiliary constraints might be non-linear. In this general form, computing solutions requires a non-convex optimization method. One prominent method for solving these types of problem is Sequential Quadratic Programming (SQP). A in-depth overview SQP methods can be found in \cite{bertsekas1999nonlinear} or \cite{wright1999numerical}. 

When using an SQP approach to solving a non-convex version of (\ref{opt:globalwords}),  newton's method is used to solve the KKT conditions of the problem \cite{wright1999numerical}. Each iteration of newton's method results in a linearly-constrained LQR problem, of which the solution provides an update to the solution of the non-convex problem. Therefore, because this procedure requires solving many constrained LQR problems, having an efficient means of computing the solutions to those subproblems is critical for an efficient solution to the non-convex problem.  
%Having an efficient solver for equality constrained QPs (as ours is) is important for the case of inequality constrained problems as well. This is because when applying SQP to inequality-constrained problems, the QP subproblems formed are also inequality-constrained. These can then be solved by means of an active-set method, which requires solving equality-constrained QPs as subproblems \cite{wright1999numerical}. However, because we are able to emphasize the properties of our control policies without incorporation of inequality constraints, we restrict our discussion to the equality-only-constrained problems. 

If the solutions of constrained LQR subproblems generated in an SQP are only used as updates in an iterative procedure for generating a trajectory, it may seem unnecessary to generate feedback policies and an \say{open-loop}approach might suffice. However, there has been much research into the advantages of \say{shooting} type methods for unconstrained variants of the nonlinear optimal control problem, such as in Differential Dynamic Programming \cite{jacobson1970differential}. These methods generate iterates by applying the \textit{open-loop} controls updates on the nonlinear system dynamics, in effect projecting the iterate onto the manifold of dynamically feasible trajectories.  \nocite{dunn1989efficient} A recent exploration into the benefits of of these type of methods \cite{giftthaler2017family} has shown that generating iterates in this way can lead to improved rate of convergence of trajectories to solutions of the non-convex problem, but sometimes suffer instabilities when the underlying system dynamics are unstable. Using the feedback control \textit{policies} to update the control signal as the nonlinear system trajectory diverges from the linear system trajectory such as in \cite{sideris2005efficient} and \cite{li2004iterative} can mitigate this instability while maintaining enhanced convergence properties. 

Because our method is highly efficient, and because it can handle arbitrary constraints without making any assumptions about linear dependence or dimension, it is an excellent candidate for use in SQP algorithms for trajectory optimization. Therefore, using our method to compute solutions to sub-problems would be no-worse than using a direct method in terms of versatility and computation-time, and the feedback policies could potentially improve convergence as discussed in \cite{giftthaler2017family} and \cite{giftthaler2017projection}. An in-depth analysis of how and when these policies can aid in convergence would be interesting, but is left for future work.

%First, if the non-convex trajectory optimization problem is subject to auxiliary constraints appearing at multiple time-steps having dimension exceeding the control dimension, methods such as in \cite{ko2007optimal}, \cite{giftthaler2017projection} and \cite{park2008lq} are not general enough to handle the linearized constraints that will appear in an LQ approximation of the problem. The method presented in \cite{sideris2011riccati} is general enough, but in highly constrained systems, the computational overhead would be such that it would require excessive computation to compute solutions. The only other sufficiently-general method we are aware of that can compute solutions to the LQ problem in comparable time is using a direct method such as LAPACK with which we compared in section \ref{sec:computation}. In that comparison, we saw that our method out-performs LAPACK on large-scale problems, and we are able to obtain useful feedback policies as a by-product of the computation. Therefore, using our method to compute solutions to sub-problems would be no-worse than using a direct method, and the feedback policies could potentially improve convergence as discussed in \cite{giftthaler2017family} and \cite{giftthaler2017projection}. An in-depth analysis of how and when these policies can aid in convergence would be interesting, but is left for future work.

\section{CONCLUSION} 
In summary, we have presented a method for computing feedback control policies for the general linearly-constrained LQR problem. The method presented has a computational complexity that scales linearly with respect to the trajectory length. We demonstrated that in practice the computation of such policies is on the order of the fastest existing methods. We also showed that the control policies generated are useful in contexts of robustly satisfying constraints, and offered perspective on the use of our method in contexts of solving general trajectory optimization problems.

\bibliographystyle{./IEEEtran} % use IEEEtran.bst style
\bibliography{root.bib}
% \bibliography{./IEEEabrv,./IEEEexample, ref.bib}

\end{document}